\documentclass[aps,prd,twocolumn,superscriptaddress]{revtex4-1}

\usepackage{longtable}
\usepackage{grffile} 
\usepackage{graphics}
\usepackage{graphicx} 
\usepackage{amsmath}    
\usepackage{hyperref}   
\usepackage{subfigure}  
\usepackage{epsfig}
\usepackage{epstopdf}
\usepackage{ulem}
\usepackage{color}

\newcommand{ \qqbar }{\mbox{$q\bar{q}$ }}

\begin{document}

\title{Impact of single string structure and multiple string interaction on strangeness production in Pb+Pb collisions at $\sqrt{S_{NN}} = 2.76$ TeV}

\author{Dai-Mei Zhou}\email{zhoudm@mail.ccnu.edu.cn}
\affiliation{Key Laboratory of Quark and Lepton Physics (MOE) and \break Institute
	of Particle Physics, Central China Normal University,\break Wuhan 430079, China}

\author{Liang Zheng}\email{zhengliang@cug.edu.cn}
\affiliation{School of Mathematics and Physics, China University of Geosciences (Wuhan),
	\break Wuhan 430074, China}

\author{Zhi-hong Song}
\affiliation{Key Laboratory of Quark and Lepton Physics (MOE) and \break Institute
	of Particle Physics, Central China Normal University,\break Wuhan 430079, China}

\author{Yu-Liang Yan}
\affiliation{China Institute of Atomic Energy, P. O. Box 275 (18), Beijing 102413, China}

\author{Gang Chen}
\affiliation{School of Mathematics and Physics, China University of Geosciences (Wuhan),
	\break Wuhan 430074, China}

\author{Xiao-Mei Li}
\affiliation{China Institute of Atomic Energy, P. O. Box 275 (18), Beijing 102413, China}

\author{Xu Cai}
\affiliation{Key Laboratory of Quark and Lepton Physics (MOE) and \break Institute
	of Particle Physics, Central China Normal University,\break Wuhan 430079, China}

\author{Ben-Hao Sa}\email{sabh@ciae.ac.cn}
\affiliation{China Institute of Atomic Energy, P. O. Box 275 (18), Beijing 102413, China}
\affiliation{Key Laboratory of Quark and Lepton Physics (MOE) and \break Institute
	of Particle Physics, Central China Normal University,\break Wuhan 430079, China}

\date{\today}

\begin{abstract}
We present a systematic study on the strange and multi-strange particles
production in Pb+Pb collisions at $\sqrt{S_{NN}} = 2.76$ TeV based on the PACIAE
model simulations. Two different mechanisms, namely the single string structure
variations and multiple string interactions, are implemented in the simulations.
These modifications give rise to an enhancement of the string tension value
involved in the Lund string fragmentation framework and generate more strange
particles in the hadronic final state. By comparing the simulation results with
the ALICE experimental data, it turns out that the inclusion of the variable
effective string tension in the PACIAE model is capable to reach an improved
agreement between theory and experiment on the strangeness production in Pb+Pb
collisions.
\end{abstract}

\maketitle

\section{Introduction}
The study of strange and multi-strange particle production in relativistic
heavy-ion collisions is an important tool to investigate the properties of the
created strongly interacting QCD system, called Quark-Gluon Plasma (QGP). The
enhancement of strangeness in heavy-ion collisions was one of the earliest
proposed signals for QGP formation. For the hadronic scattering in vacuum, the
strangeness content in the created particles is much smaller compared to the
lighter components due to the mass suppression effect in the particle
productions. The mass of the strange quark is close to the deconfinement
temperature needed to create the QGP matter, allowing for the thermal production
of strange quarks and also favoring the formation of multi-strange
hadrons~\cite{Rafelski, Koch}. As the lifetime of the QGP is estimated to be
long enough for the full equilibration of strange quarks, a higher abundance of
strangeness production per participant pair is expected in heavy-ion collisions
than what is seen in proton-proton interactions. ALICE Collaboration reported a
series of surprising strangeness enhancement observations in small collision
systems like pp and pA~\cite{Alicenature,Acharya:2019kyh,Adam:2015vsf}. They
measured the yield ratios of multi-strange hadrons to charged pions as a
function of multiplicity. These ratios are found to grow rapidly with
multiplicity in pp collisions, and reach the values close to ones in PbPb
collisions at full equilibrium in high multiplicity pPb collisions. It is
interesting to notice that the strangeness-to-pion ratios do not depend on the
energy and size of collision system.

Different models are trying to interpret this universal strangeness-to-pion
ratio as a function of the event multiplicity. The THERMUS statistical
hadronization model suggest the suppression of strange hadron production in pp
may come from the explicit conservation of strangeness quantum number based on
the canonical approach~\cite{Vislavicius, Alice-thermal}. The EPOS model~
\cite{Pierog} assumes the QGP matter is partly formed in the pp collisions
treating the interactions based on a core-corona approach. The DIPSY
model~\cite{Bierlich:2014xba} employs the color ropes
mechanism~\cite{Biro:1984cf,Andersson:1991er} taking into account the color
interactions between overlapping strings and usually can reasonably describe the
hierarchy of strangeness production. Other interesting extensions to the Lund
string fragmentation model considering the thermodynamic features of strings in
a dense environment~\cite{Fischer:2016zzs} are implemented in
PYTHIA8~\cite{Sjostrand:2014zea} for data comparison as well. While various
models can describe some key features of the data, the fundamental origin of
enhanced strangeness production is still largely unknown. It suggests further
developments are needed to reach a complete microscopic understanding of
strangeness production in small systems and to identify whether such mechanisms
would contribute to the observed effects in heavy-ion collisions.

In the Ref.~\cite{paciae-strangness-pp}, we introduced an effective string
tension stemming from the single string structure and multiple string
interaction to reduce the strange quark suppression in pp collisions based on
the tunneling probability in Lund string fragmentation regime. These two
mechanisms are implemented in the PACIAE Monte Carlo event generator~\cite{sa1}
to study the strange hadron production in $\sqrt{s}=$7 TeV pp collisions.
We find that the inclusion of variable effective string tension yields an
improved agreement between theory and experiment, especially for the recently
observed multiplicity dependence of strangeness enhancement in pp collisions.
This approach provides us a new method to understand the microscopic picture
of the novel high multiplicity pp events collected at the LHC in the string
fragmentation framework. In this work, we will extend this framework to
describe heavy-ion collisions and study the strange and multi-strange particle
productions in Pb+Pb collisions at $\sqrt{S_{NN}} = 2.76$ TeV.

\section{Model setup}
The PACIAE model is based on parton initial states generated by PYTHIA
convoluted with the nuclear geometry implementing via the Glauber model
~\cite{soj1}. The partonic rescattering is introduced after the creation of
parton initial condition while the hadronic rescattering may happen after the
hadronization of QCD matter. Thus, the PACIAE model can be employed to simulate
a wide range of collision systems from nucleon-nucleon collision system to
nuclear-nuclear collision system. We give more detailed information about PACIAE
model in \cite{sa1}.

In this paper we extend the effective string tension approach to the
description of nuclear-nuclear collisions by assuming the parametrization for
single string structure is the same as that in pp collisions. Therefore the
parameterized effective string tension responsible for the single string
structure is given as:
\begin{equation}
\kappa_{eff}^{s}=\kappa_0 (1-\xi)^{-a},
\label{eqn:single_kappa}
\end{equation}
where $\kappa_0$ is the pure \qqbar string tension usually set to 1 GeV/fm,
$a$ is a parameter to be tuned with experimental data while $\xi$ can be
parameterized as:
\begin{equation}
\xi = \frac{\ln(\frac{k_{\perp max}^2}{s_0})}{\ln(\frac{s}{s_0})+\sum_{j=gluon}\ln(\frac{k_{\perp j}^2}{s_0})},
\end{equation}
with $k_{\perp}$ being the transverse momentum of the gluons inside a dipole
string. $\sqrt{s}$ and $\sqrt{s_0}$ give the mass of the string system and a
parameter related to the typical hadron mass, respectively. The quantity $\xi$
quantifies the difference of a gluon wrinkled string compared to a pure \qqbar
string, assuming the fractal structure of a string object is dominated by the
hardest gluon on the string. $(1-\xi)^{-1}$ in Eq.~\ref{eqn:single_kappa} thus
describes the multiplicity enhancement factor of the hardest gluon to the rest
of the string component and can be related to the string tension with a scaling
formula. The value of this string tension changes on a string-by-string basis in
the current implementation and takes the string-wise fluctuations into
consideration.

The multiple string interaction effects raised from the correlation of strings
overlapped in a limited transverse space are parameterized in a similar way
of the close-packing strings~\cite{Fischer:2016zzs} in pp collisions as
follows:
\begin{equation}
\kappa_{eff}^{m}=\kappa_0 (1+\frac{n_{MPI}-1}{1+p^{2}_{T\ ref}/p^{2}_0})^{r},
\label{eqn:global_kappa-p}
\end{equation}
In heavy ion collisions, the multiple string interaction
effects will be stronger than pp collisions, due to the enhanced string
density generated with multiple nucleon-nucleon collisions. Thus, we extend
Eq.~\ref{eqn:global_kappa-p} to
\begin{equation}
\kappa_{eff}^{m}=\kappa_0 (1+\frac{\frac{N_{coll}}{N_{part}}n_{MPI}-1}{1+p^{2}_{T\ ref}/p^{2}_0})^{r},
\label{eqn:global_kappa-pb}
\end{equation}
in which $n_{MPI}$ indicates the number of multiple parton interactions in one
nucleon-nucleon collision. $N_{coll}$ is the number of binary collisions and
$N_{part}$ is the number of participant nucleons in a nuclear-nuclear
collision. Other parameters are chosen to be $p^{2}_{T\ ref}/p^{2}_0=1$,
$r=0.2$, the same as in pp collisions ~\cite{paciae-strangness-pp}.

Additionally, the string-wise single string structure and event-wise multiple
string interaction effects can be combined together by replacing the
$\kappa_0$ in Eq.~\ref{eqn:global_kappa-pb} with $\kappa_{eff}^{s}$. Assuming
$\kappa_0= 1$ GeV/fm, one can derive a combined effective string tension as a
product of the two:
\begin{eqnarray}
\kappa^{s+m}_{eff} = & \kappa^{s}_{eff} (1+\frac{\frac{N_{coll}}{N_{part}}n_{MPI}-1}{1+p^{2}_{T\ ref}/p^{2}_0})^{r} \nonumber \\
  = & \kappa^{s}_{eff}\times\kappa^{m}_{eff}.
 \label{eqn:couple_kappa-pb}
\end{eqnarray}

\begin{figure*}[hbt!]

\includegraphics[width=0.55\textwidth]{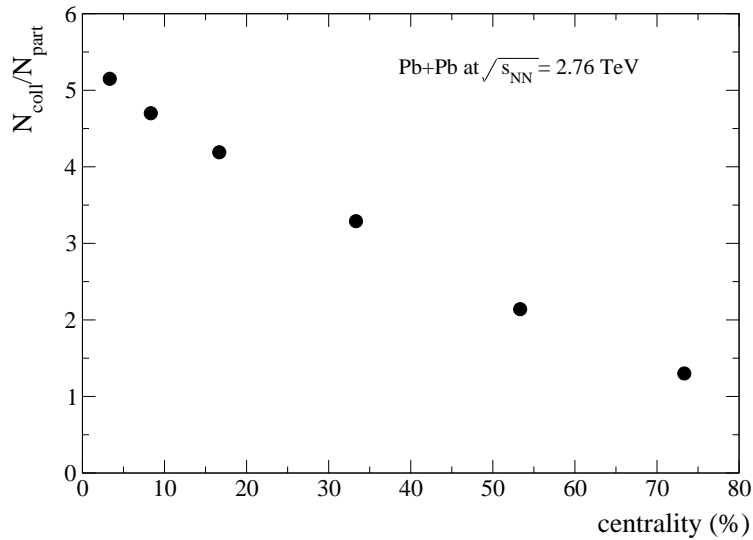}
\caption{$N_{coll}/N_{part}$ varying with the collision centrality}.
\label{ratincollnpart}
\end{figure*}


The factor of $N_{coll}/N_{part}$ introduced in Eq.(\ref{eqn:couple_kappa-pb})
amplifies the multiple string interaction effects and delivers the variational
string tension at high multiplicity for nuclear-nuclear collisions.
Figure.~\ref{ratincollnpart} shows the ratio of $N_{coll}$ to $N_{part}$ varying
with collision centrality from the PACIAE calculations. This term accounts
for the major system size dependence effect in our effective string tension
approach.

In the Lund string fragmentation regime, the following parameters are expected
to evolve according to the change of the effective string tension:
 \begin{itemize}
 \item PARJ(1) is the suppression of diquark-antidiquark pair production
compared to the quark-antiquark production,
 \item PARJ(2) is the suppression of $s$ quark pair production compared to
the $u$ or $d$ pair production,
 \item PARJ(3) is the extra suppression of strange diquark production
compared to the normal suppression of strange quark,
 \item PARJ(21) (denoted by $\sigma$) corresponds to the width $\sigma$ in the
Gaussian $p_x$ and $p_y$ transverse momentum distributions for primary
hadrons.
 \end{itemize}

These Lund strng fragmentation parameters can be related to the effective
string tension through a scaling function implied by the tunneling probability:
 \begin{equation}
 \lambda_2=\lambda_1^{\frac{\kappa^{eff}_1}{\kappa^{eff}_2}},
 \label{lamd}
 \end{equation}
where $\lambda$ refers to PARJ(1) or PARJ(2) or PARJ(3) etc. In the above
equation, $\kappa^{eff}_1$ represents the string tension in vacuum (usually
assigned with $\kappa^{eff}_1$=1 GeV/fm) and $\lambda_1$ shows the corresponding
fragmentation parameter values. $\kappa^{eff}_2$
gives the effective string tension taking into account the string tension variation
impacts. The fragmentation parameter $\lambda_2$ will be enlarged when the effective
string tension $\kappa^{eff}_2$ becomes greater than $\kappa^{eff}_1$. Similarly,
the fragmentation $p_T$ width PARJ(21) varies with the effective string tension as
 \begin{equation}
 \sigma_2=\sigma_1(\frac{\kappa^{eff}_2}{\kappa^{eff}_1})^{1/2}.
 \end{equation}

In the rest of the work, we will focus on the comparison of our PACIAE
calculations to the ALICE PbPb 2.76 TeV data and study the variational string
tension effects on the centrality dependence of strange hadron productions. In
order to describe the mid-rapidity charged particle yields
$\frac{dN_i}{dy}|_{|y|<0.5}$ of ALICE experimental
results~\cite{ALICE:2013prc88-044910} by PACIAE model, three parameters are
tuned： PARP(31)=2.7($K$ factor) for the hard scattering cross sections,
PARJ(41)=1.7($\alpha$)and PARJ(42)=0.09($\beta$) in the LUND string
fragmentation function. PACIAE model gives mid-rapidity particle yields
$\frac{dN_i}{dy}|_{|y|<0.5}$ for $\pi^{+}=722.5$ and $\pi^{-}=722.2$ for $0-5\%$
centrality case, consistent with the corresponding ALICE data $\pi^{+}=733 \pm
54$ and $\pi^{-}=732 \pm 52$. These parameters were fixed for the simulation of
other event centralities. The other fragmentation parameters modified by the
effective string tension are determined with the reference values PARJ(1)=0.1,
PARJ(2)=0.3, PARJ(3)=0.4 and PARJ(21)=0.85 in the vacuum.

\begin{figure*}[hbt!]
	\centering
	\subfigure{
		\includegraphics[width=0.45\textwidth]{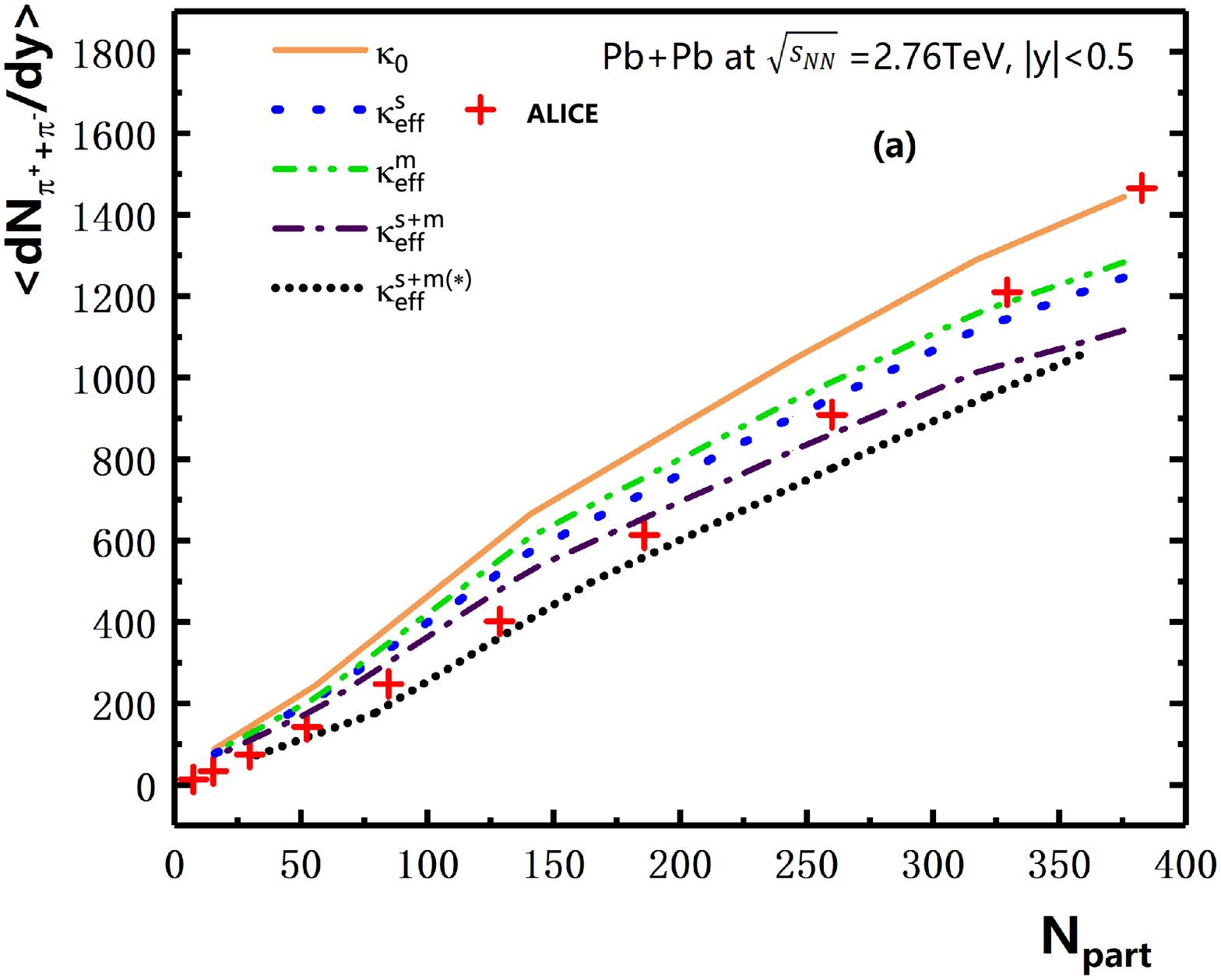}
		\label{fig:pion-npart}
	}
	\subfigure{
		\includegraphics[width=0.45\textwidth]{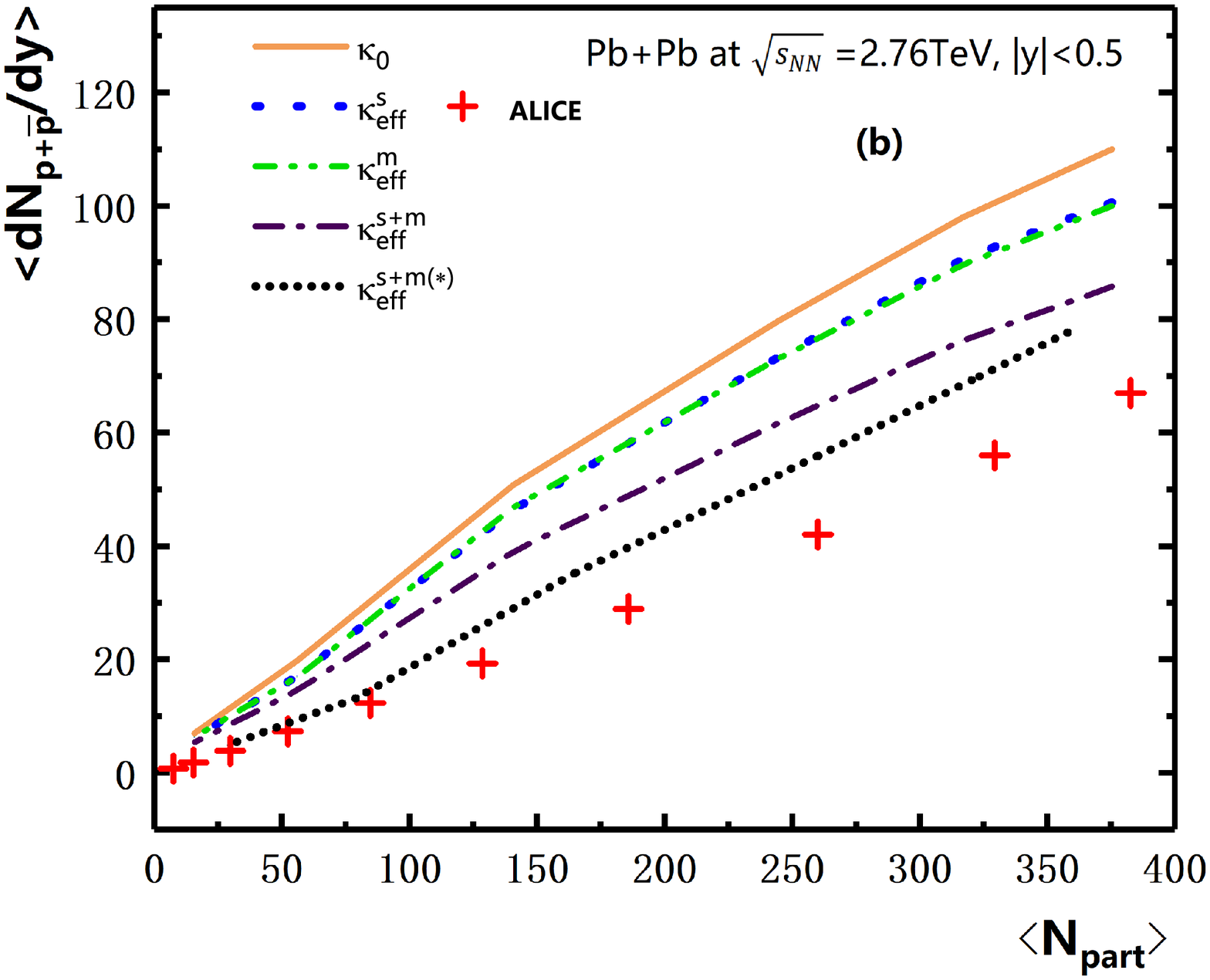}
		\label{fig:proton-npart}
	}
	\caption{(Color online) Pion (a) and proton (b) midrapidity particle
yields $\frac{dN_i}{dy}|_{|y|<0.5}$ varying with  $\langle N_{part} \rangle$
in PbPb collisions at $\sqrt{S_{NN}}=2.76$ TeV for different string environments:
constant string tension ($\kappa_0$, solid line), single string structure
($\kappa_{eff}^{s}$, dashed line), multiple string interaction
($\kappa_{eff}^{m}$, dash-dot-dotted line), single+multiple string environment
($\kappa_{eff}^{s+m}$, dash-dotted line), as well as the single+multiple with
a zero approximation of PARJ(2) and PARJ(3) enlarged to 1.5 times of
the corresponding default values ($\kappa_{eff}^{s+m(*)}$, dotted line).
The experimental data are denoted as crosses and taken
from~\cite{ALICE:2013prc88-044910}. }
	\label{fig:pion_proton_npart}
\end{figure*}

\section{Results}
To investigate the influence of different effective string tensions,
on the strangeness production in PbPb collisions at $\sqrt{S_{NN}}=2.76$ TeV,
we first perform an examination on the mean participant number dependence of
the charged pion and proton yields in Fig.~\ref{fig:pion_proton_npart}. Model
predictions with the default string tension $\kappa_{0}$, single string
structure $\kappa_{eff}^{s}$, multiple string interactions $\kappa_{eff}^{m}$,
the combination of $\kappa_{eff}^{s+m}$, and the $\kappa_{eff}^{s+m(*)}$ with
$1.5$ times enlarged default parameters of PARJ(2) and PARJ(3) are shown
as the solid, dashed, dash-dot-dotted, dash-dotted, and the dotted curves,
respectively. ALICE data are indicated by the red markers~\cite{ALICE:2013prc88-044910}.
The results indicate scenarios with larger string tension, in which strings become
more difficult to break up, generally produce less
pions and protons. The impacts on pion or proton multiplicity can be as large as
30\% varying from $\kappa_0$ to $\kappa_{eff}^{s+m}$.

\begin{figure*}[hbt!]
	\centering
	\subfigure{
		\includegraphics[width=0.45\textwidth]{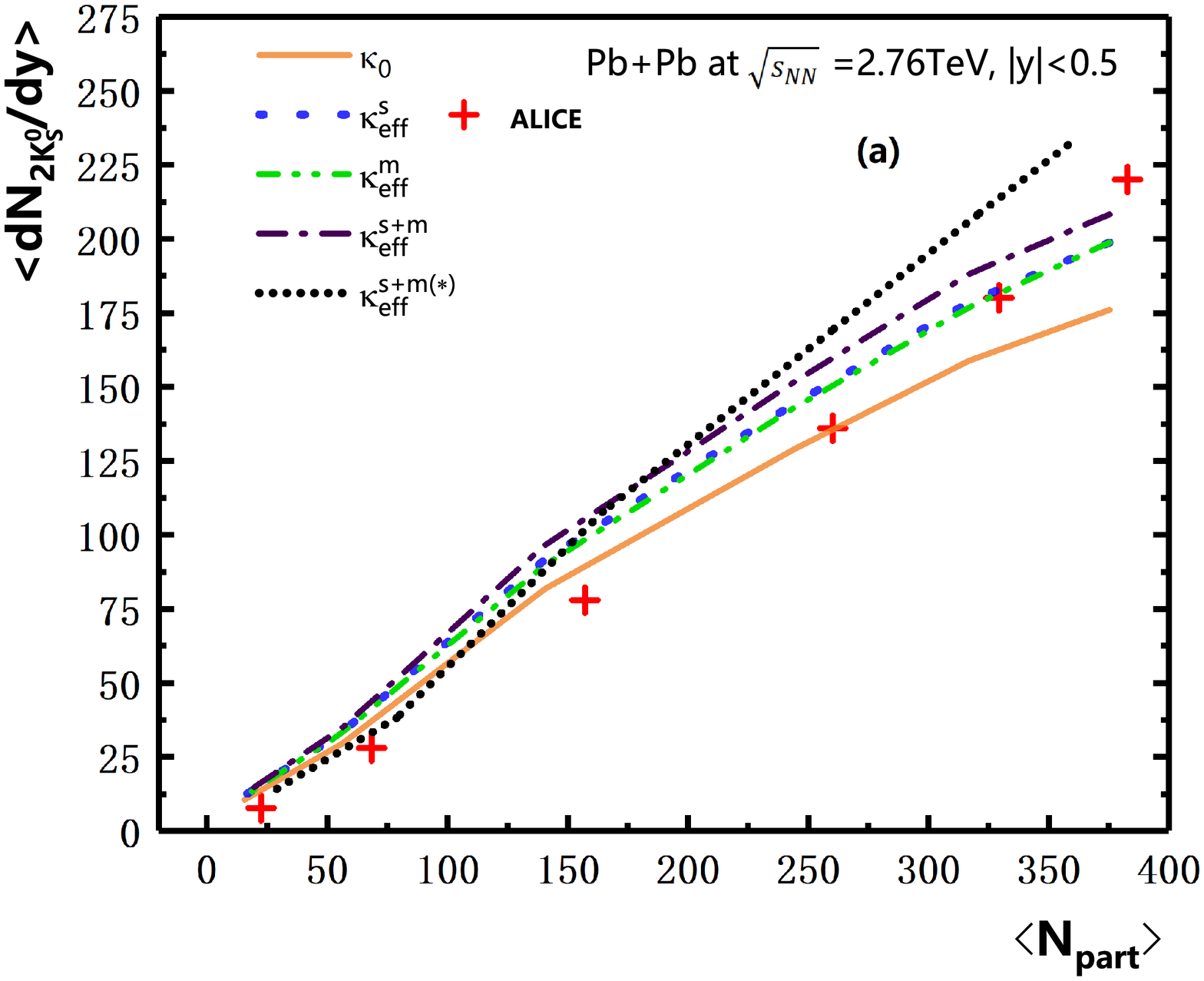}
		\label{fig:KS_npart}
	}
	\subfigure{
		\includegraphics[width=0.45\textwidth]{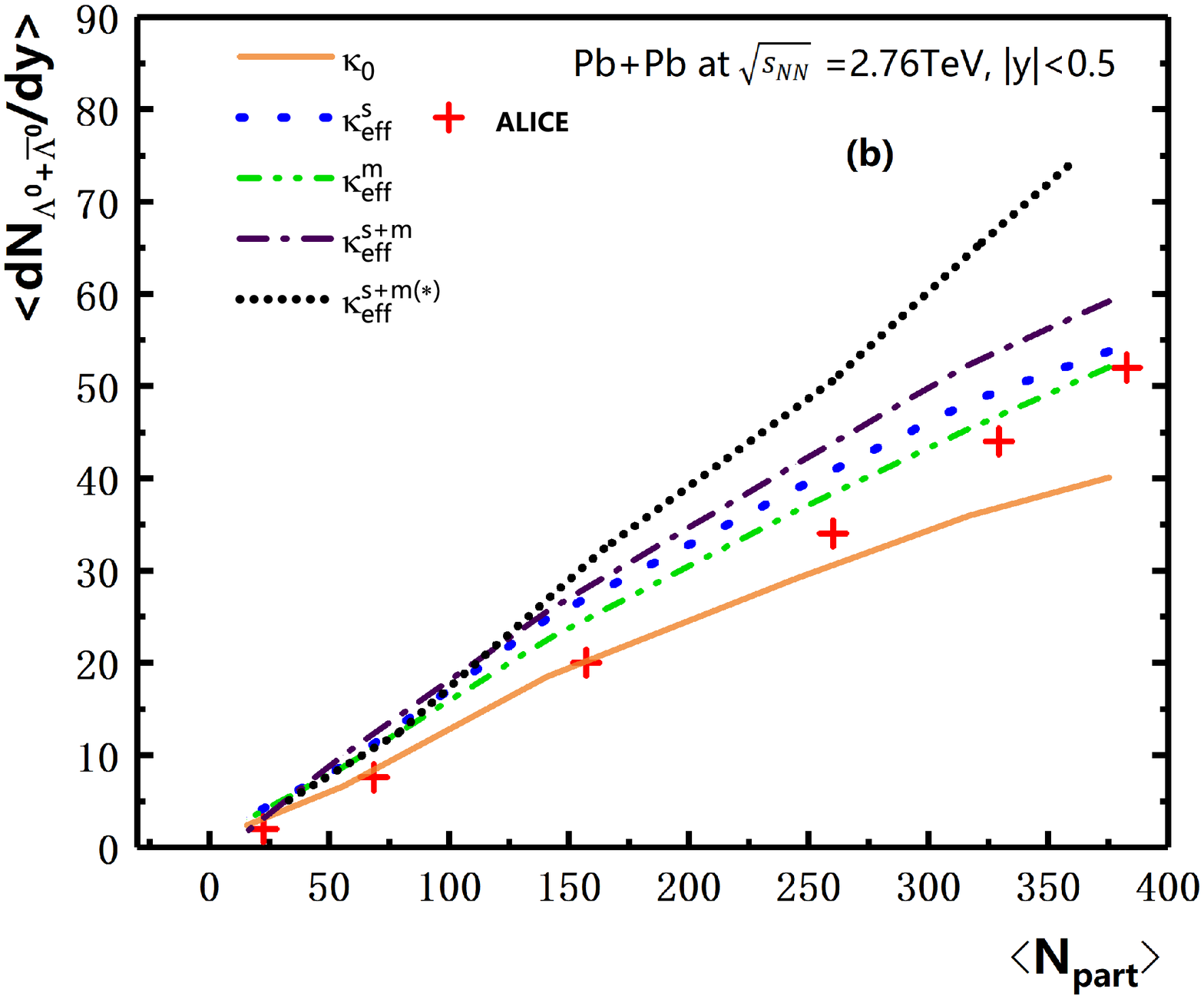}
		\label{fig:lamda_npart}
	}
	\subfigure{
		\includegraphics[width=0.45\textwidth]{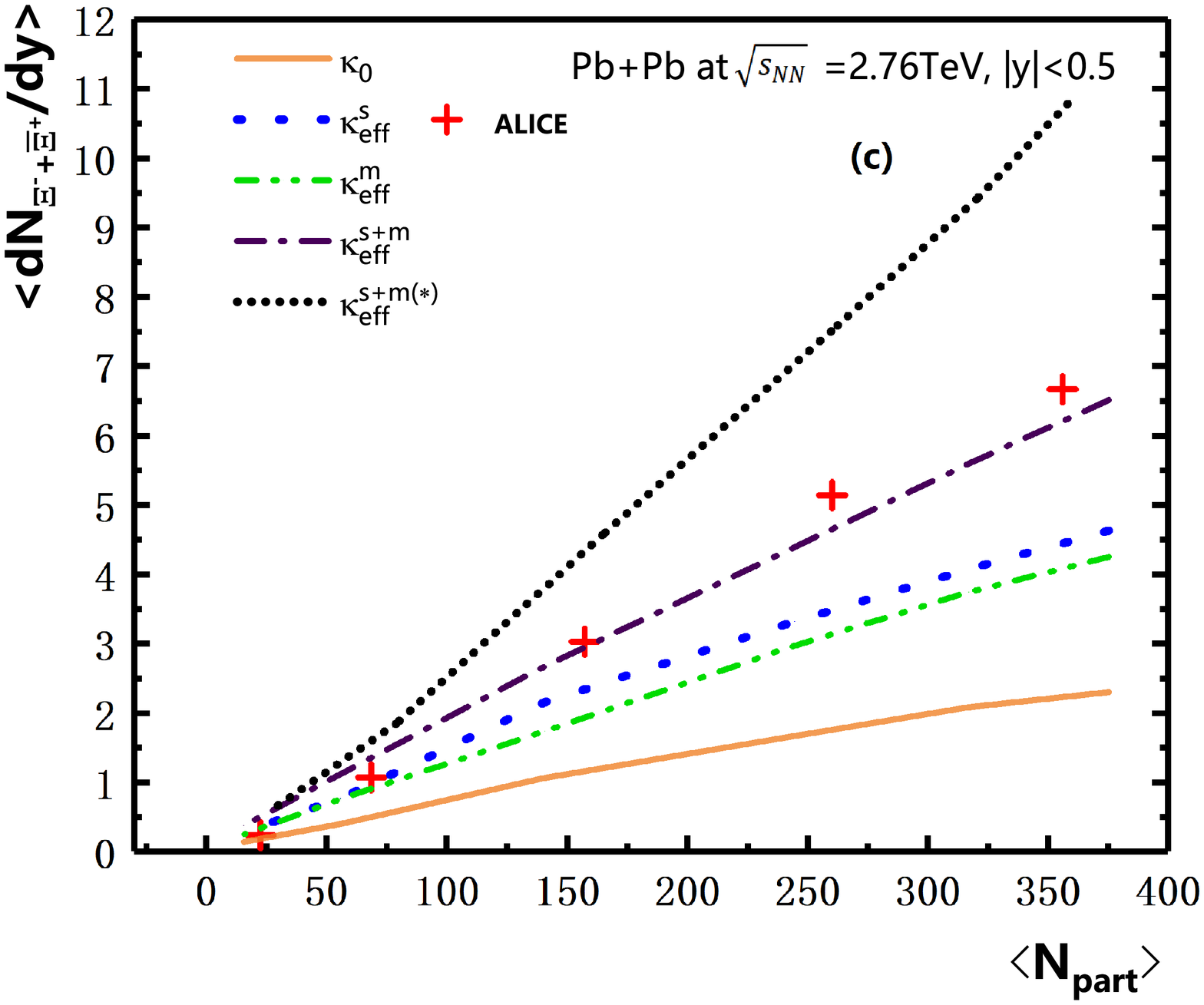}
		\label{fig:Xi_npart}
	}
	\subfigure{
		\includegraphics[width=0.45\textwidth]{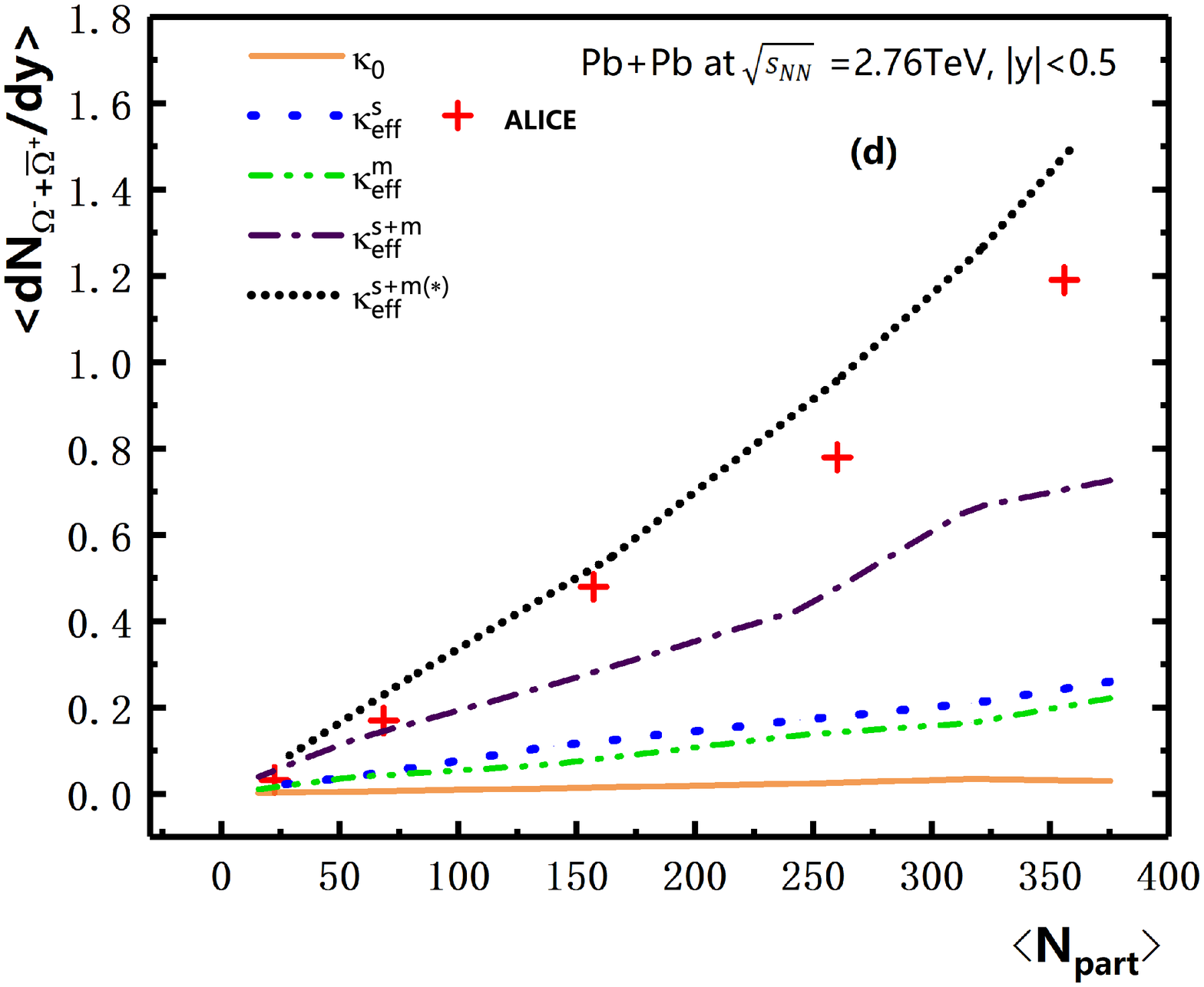}
		\label{fig:Omega_npart}
	}
	\caption{(Color online) Integrated strange and multi-strange
mid-rapidity yields, $dN/dy$, varying with the $\langle N_{part} \rangle$ in
the PbPb collisions at $\sqrt{S_{NN}}=2.76$ TeV: $K_S^{0}$ (a), $\Lambda$ (b),
$\Xi$ (c) and $\Omega$ (d).  Five different string environments are shown:
$\kappa_0$, solid curve; $\kappa_{eff}^{s}$, dashed curve; $\kappa_{eff}^{m}$,
dash-dot-dotted curve; $\kappa_{eff}^{s+m}$, dash-dotted line; and
$\kappa_{eff}^{s+m(*)}$, dotted curve. The experimental data for $K_S^{0}$
and $\Lambda$ are taken from~\cite{ALICE:2013PRL111}, for $\Xi$ and $\Omega$
are taken from~\cite{ALICE:2014PLB728}. }
	\label{fig:strange_integral_npart}
\end{figure*}

In Fig.~\ref{fig:strange_integral_npart}, we show a comparison of the simulation
results for particles with different strangeness number dependent on $\langle
N_{part} \rangle$ to ALICE data. All the strange particle productions are found
to be enhanced with the inclusion of increased effective string tension. As can
be seen in Fig.~\ref{fig:KS_npart}, $K_S^{0}$ calculation with $\kappa_{0}$ is
consistent with data in peripheral collisions, while larger effective string
tension is needed especially for the most central collisions, comparing with the
ALICE measurement~\cite{ALICE:2013PRL111}. The impact of effective string
tension on the strange baryon production is much more pronounced than that on
$K_S^{0}$. In the strange baryon comparisons, the slope change due to effective
string tension shows a clear hierarchy depending on the strangeness number. The
change of the slope from the minimum effective string tension ($\kappa_0$) to
the maximum one ($\kappa_{eff}^{s+m(*)}$) in $\Lambda$ production is the
smallest, while the $\Omega$ slope changes most dramatically from $\kappa_0$ to
$\kappa_{eff}^{s+m(*)}$ environments. The multiple string interaction
($\kappa_{eff}^{m}$) seems to be important to fit the $\Lambda$ result in these
comparisons~\cite{ALICE:2013PRL111}. The $\Xi$ result~\cite{ALICE:2014PLB728}
can be described reasonably with the calculation based on the
$\kappa_{eff}^{s+m}$ assumption. However, for $\Omega$
productions~\cite{ALICE:2014PLB728}, one will need to increase the reference
values of PARJ(2) and PARJ(3) by a factor of 1.5 to describe the data when both
single string structure variation and multiple string interaction effects on
($\kappa_{eff}^{s+m(*)}$).

\begin{figure*}[hbt!]
	\centering
	\subfigure{
		\includegraphics[width=0.45\textwidth]{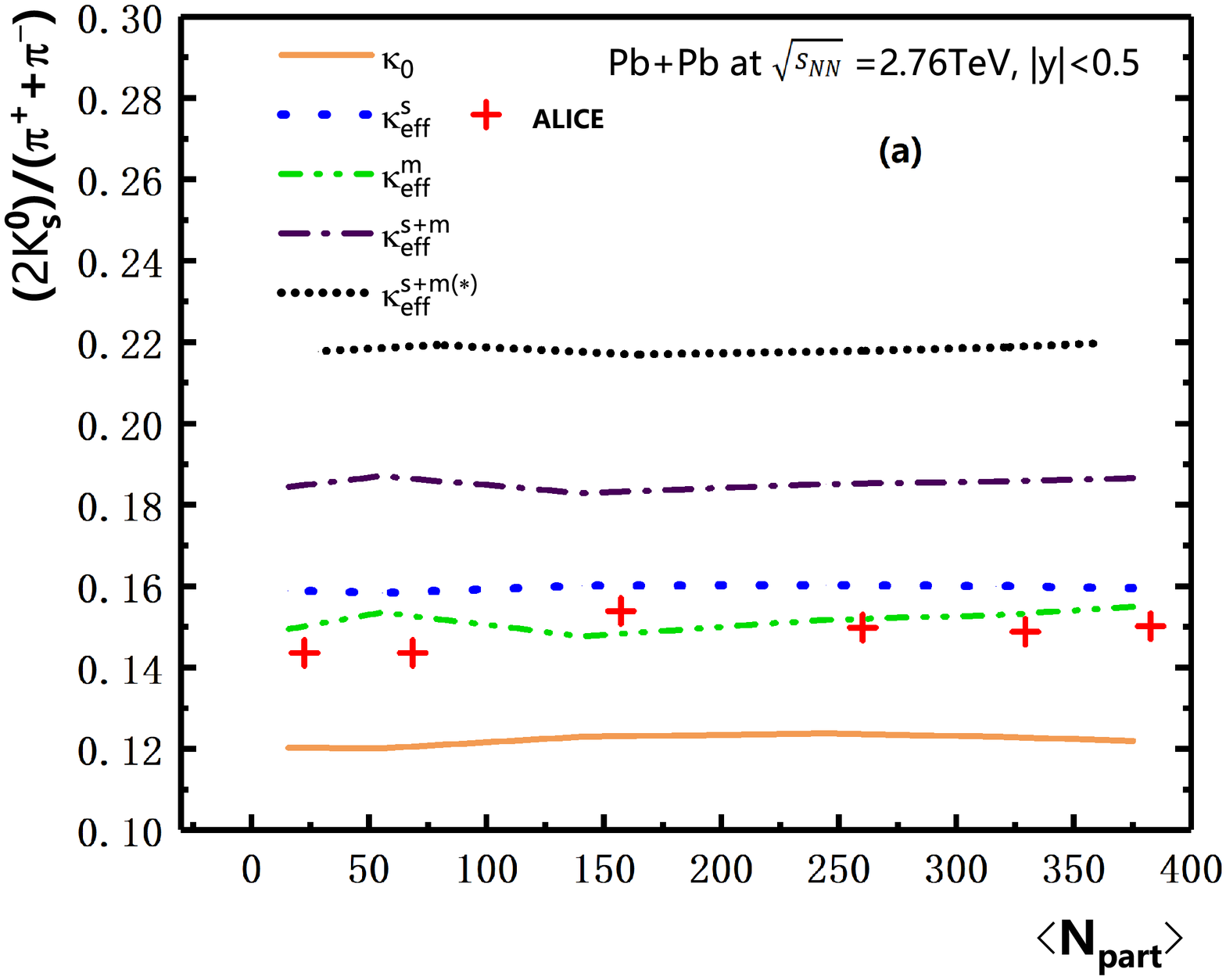}
		\label{fig:KSOverPi_npart}
	}
	\subfigure{
		\includegraphics[width=0.45\textwidth]{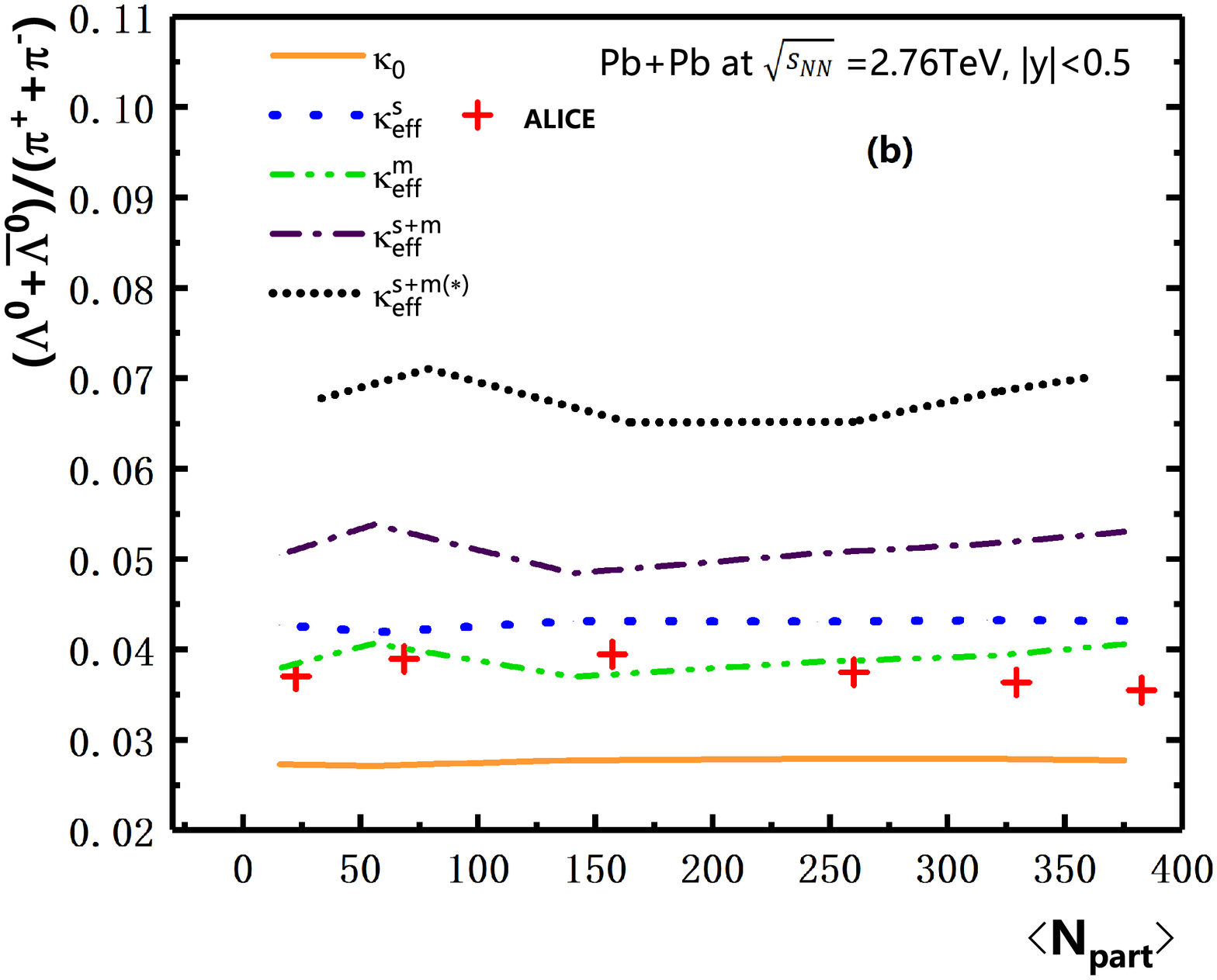}
		\label{fig:LOverPi_npart}
	}
	\subfigure{
		\includegraphics[width=0.45\textwidth]{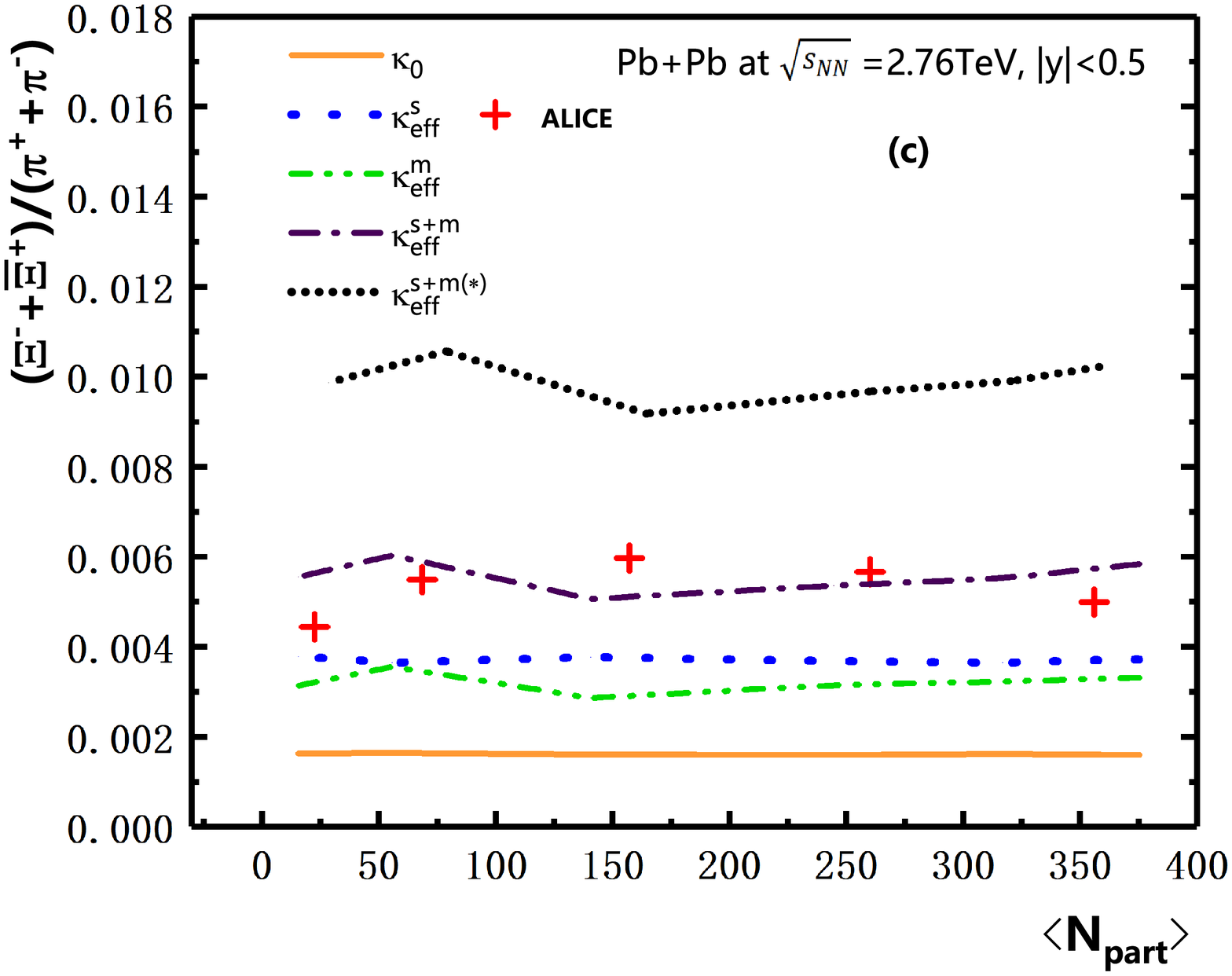}
		\label{fig:XiOverPi_npart}
	}
	\subfigure{
		\includegraphics[width=0.45\textwidth]{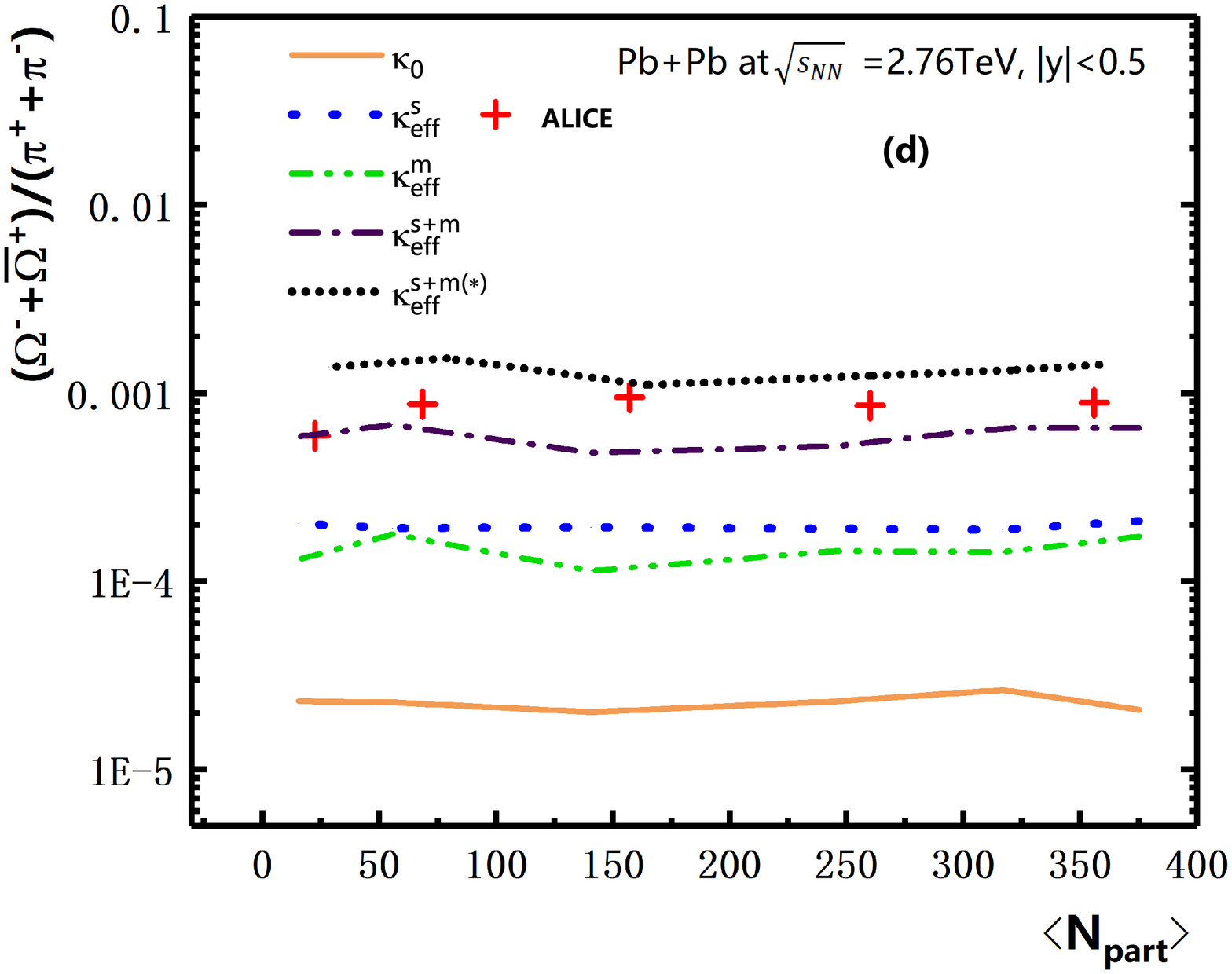}
		\label{fig:OmegaOverPi_npart}
	}
	\caption{(Color online) Ratio of strange and multi-strange particles
to pion varying with the $\langle N_{part} \rangle$ in Pb+Pb collisions at
$\sqrt{S_{NN}}=2.76$ TeV.  Five different string environment are shown: $\kappa_0$,
solid curve; $\kappa_{eff}^{s}$, dashed curve; $\kappa_{eff}^{m}$,
dash-dot-dotted curve; $\kappa_{eff}^{s+m}$, dash-dotted curve; and
$\kappa_{eff}^{s+m(*)}$, dotted curve. The ALICE data are denoted as crosses
and taken from~\cite{ALICE:2013prc88-044910,ALICE:2013PRL111,ALICE:2014PLB728}  }
	\label{fig:strangeOverPion_npart}
\end{figure*}

Aside from examining the integrated yield of particles as a function of the
$\langle N_{part} \rangle$, it is also of great interest to understand the
relative production of strange particles.
Figure.~\ref{fig:strangeOverPion_npart} shows strange particle relative to $\pi$
production for $K_S^{0}$, $\Lambda$, $\Xi$ and $\Omega$. It is obvious to see
these ratios strongly depend on the choice of the string tension assumptions.
For the situation with large effective string tension, the strange to pion
ratios are significantly enhanced. The magnitude of enhancement is related to
the strangeness number. The factor of enhancement is found to be larger for the
multi-strange baryons. On the other hand, it is also a bit surprising to see the
strange to pion ratios are almost independent of the $\langle N_{part} \rangle$
in each case. This behavior can be understood with the scaling feature of our
two string tension variation mechanism. For the single string structure, it is
mostly energy dependent and has no strong correlation with the $\langle N_{part}
\rangle$. For the multiple string interaction, the effective string tension
changes rather slowly with $N_{part}$ as $r$ is 0.2 in
Eq.~\ref{eqn:global_kappa-pb} leading to a mild growth on the system size. By
combining these two effects, the magnitude of effective string tension may
change significantly with different string tension assumptions, but does not
rely on the event centrality. The weak $N_{part}$ dependence of multi-strange to
pion ratios in the heavy ion collision system provided by these model
implementations are qualitatively confirmed by the ALICE data. This comparison
implies the effective string tension approach can be an essential
phenomenological tool to understand the origin of strangeness enhancement
effect.

\section{Discussions and Conclusions}
We provide a systematic study on the strange particle productions in high
energy PbPb collisions based on the variational effective string tension
approach in PACIAE. In the comparison of integrated particle yields, the
effects of string tension variations are found to be important to describe the
$\langle N_{part} \rangle$ dependence especially for the multi-strange
hadrons. The slope change due to effective string tension shows a clear
hierarchy depending on the strangeness number.

The strange particles relative to $\pi$ production are studied as well. PACIAE
model simulations elucidate that for different string tensions, the ratios
are generally larger with stronger effective string tension. However, for each
string tension assumption, the strange to pion ratios are rather flat across
different event centralities, which is consistent with the observations in
the ALICE data. Our findings in this work suggest the effective string tension
approach implemented in PACIAE can provide the qualitative feature of the
enhanced strange particle production, which has been considered to be a
defining signal for QGP creation in heavy ion collisions for a long time. The
implementation of this framework in PACIAE will pave the way for a new
perspective to understand the strangeness enhancement effect observed in
nuclear collisions.

\begin{acknowledgments}
This work was supported by the National Natural Science Foundation of China (11775094, 11905188, 11775313), the Continuous Basic Scientific Research Project (No.WDJC-2019-16), National Key Research and Development Project (2018YFE0104800) and by the 111 project of the foreign expert bureau of China.
\end{acknowledgments}


\end{document}